\documentclass[aps,prb,reprint,superscriptaddress]{revtex4-1}
\usepackage[dvips]{graphicx}
\usepackage{amsmath}
\usepackage{color}

\begin{document}

\title{Exciton-exciton annihilation in MoSe$_2$ monolayers}

\author{Nardeep Kumar}
\affiliation{Department of Physics and Astronomy, The University of Kansas, Lawrence, Kansas 66045, USA}

\author{Qiannan Cui}
\affiliation{Department of Physics and Astronomy, The University of Kansas, Lawrence, Kansas 66045, USA}

\author{Frank Ceballos}
\affiliation{Department of Physics and Astronomy, The University of Kansas, Lawrence, Kansas 66045, USA}

\author{Dawei He}
\affiliation{Key Laboratory of Luminescence and Optical Information, Ministry of Education, Institute of Optoelectronic Technology, Beijing Jiaotong University, Beijing 100044, China}

\author{Yongsheng Wang}
\email{yshwang@bjtu.edu.cn}
\affiliation{Key Laboratory of Luminescence and Optical Information, Ministry of Education, Institute of Optoelectronic Technology, Beijing Jiaotong University, Beijing 100044, China}

\author{Hui Zhao}
\email{huizhao@ku.edu}
\affiliation{Department of Physics and Astronomy, The University of Kansas, Lawrence, Kansas 66045, USA}

\date{\today}

\begin{abstract}
We investigate the excitonic dynamics in MoSe$_2$ monolayer and bulk samples by femtosecond transient absorption microscopy. Excitons are resonantly injected by a 750-nm and 100-fs laser pulse, and are detected by a probe pulse tuned in the range of 790 - 820 nm.  We observe a strong density-dependent initial decay of the exciton population in monolayers, which can be well described by the exciton-exciton annihilation. Such a feature is not observed in the bulk under comparable conditions. We also observe the saturated absorption induced by exciton phase-space filling in both monolayers and the bulk, which indicates their potential applications as saturable absorbers.

\end{abstract}

\maketitle

Layered materials in which atomic sheets are stacked together by the weak van der Waals force can be used to fabricate two-dimensional (2D) systems, which can have exotic properties that are very different from their bulk counterparts. They represent a new approach to develop nanomaterials. Since 2004, most efforts have been focused on graphene.\cite{s306666,s3121191,n457706,nn4712} More recently, however, other layered materials have drawn considerable attentions.\cite{rpp74082501,nn7699,acsnano72898} For example, atomically thin semiconducting transition metal dichalcogenides, MX$_2$ (M=Mo, W; X=S, Se, Te), have shown several interesting properties, such as transition to a direct bandgap in monolayers,\cite{l105136805,nl101271,b84045409,b85205302,l111106801} valley-selective optical coupling,\cite{l108196802,nn7490,nn7494,nn8634,nc3887,nc42053,np9149} and large nonlinear optical responses.\cite{sr31608,b87161403,b87201401,nl133329} Various applications of monolayer MX$_2$ have also been developed, including field-effect transistors,\cite{nn6147,nl113768,nl121136,b85115317} integrated circuits, \cite{acsnano59934,nl124674} phototransistors,\cite{nl123695} chemical sensors,\cite{nl13668} and light-emitting diodes.\cite{nl131416} 

In these atomically thin 2D structures, the exciton binding energies\cite{b86115409,nm12207,nc41474,b86241201} are much larger than semiconductor quantum wells - the previously extensively studied quasi-2D systems. Hence, they provide a new platform to study excitons in confined systems. Since the optical properties of these systems are dominated by excitons even at room temperature, for various applications, it is important to understand their excitonic dynamics. Here we report an ultrafast optical study of the excitonic dynamics in MoSe$_2$ monolayers. 

So far, most studies on MX$_2$ have focused on one member of this family, namely MoS$_2$. Other members have similar lattice structures as MoS$_2$, but possess different properties, such as the sizes of the bandgap and the strengths of spin-orbital coupling.\cite{nn7699} Hence, they can potentially be used to complement MoS$_2$ in some applications. More importantly, it is possible to use various types of atomic layers as building blocks to assemble multilayer structures and even three-dimensional crystals to achieve desired properties.\cite{n499419} Therefore, understanding the basic properties of these building blocks is essential. Recently, strong exciton\cite{nl125576} and trion\cite{nc41474} photoluminescence has been observed in monolayers of MoSe$_2$ using time-integrated measurements. The high temporal and spatial resolution of our transient absorption microscopy measurements allows us to directly study dynamics of excitons in MoSe$_2$ monolayers. We observe efficient exciton-exciton annihilation at high exciton densities, which reveals the strong interaction between excitons in this strongly confined system. Similar measurements performed on a bulk sample indicate that this process is absent in bulk. 

MoSe$_2$ monolayer samples are fabricated by mechanical exfoliation with an adhesive tape from bulk crystals (2D Semiconductors). By depositing flakes of MoSe$_2$ on silicon substrates with either a 90-nm or a 280-nm SiO$_2$ layer, we can identify large flakes of MoSe$_2$ monolayers with an optical microscope, by utilizing optical contrasts enhanced by the multilayer substrate.\cite{nanotechnology22125706,apl96213116} Photoluminescence and Raman spectroscopy are also performed to confirm the thickness of the flakes studied. All the measurements were performed in ambient condition and no signs of sample degradation were observed during the entire study. 

In the transient absorption measurements, an 80-MHz mode-locked Ti:sapphire laser is used to generate 100-fs pulses with a central wavelength in the range of 790 - 820 nm. The majority of this beam is used to pump an optical parametric oscillator, which has a signal output of 1500 nm. To obtain the pump pulse for the measurement, a beta barium borate (BBO) crystal is used to generate the second harmonic of this beam, with a wavelength of 750 nm. Tuned to the high-energy edge of the A-exciton resonance, the pump pulse injects excitons by resonant excitation. The injected excitons are probed by a 100-fs pulse with a different wavelength in the low-energy side of the resonance. It is obtained directly from the Ti:sapphire laser. The reflected probe is directed to a photodetector, which output is measured by a lock-in amplifier. Balanced detection is used to improve the signal-to-noise ratio of the system.\cite{l106107205} By using a microscope objective lens, we tightly focus both the pump and the probe pulses to spot sizes of about 1 $\mu$m, which is several times smaller than the dimensions of the flakes studied. The pump and the probe spots are overlapped, and are located near the center of the flakes in all the measurements. 

\begin{figure}
 \includegraphics[width=8.5cm]{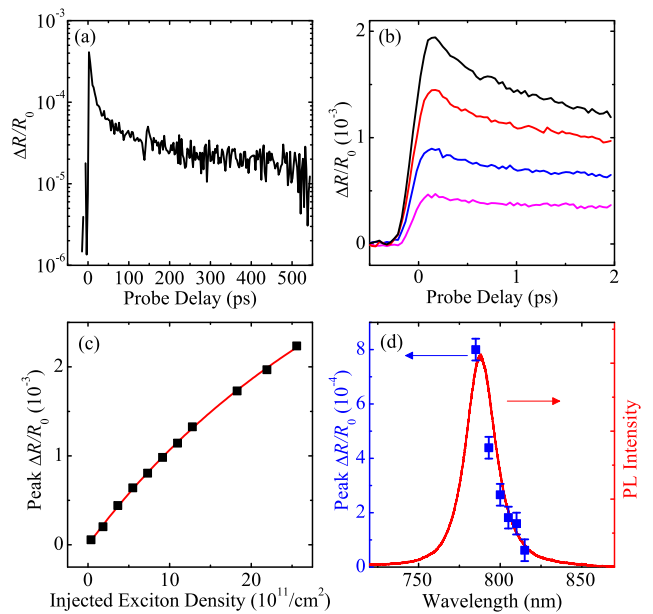}
 \caption{(a) Differential reflection of a MoSe$_2$ monolayer measured with a probe wavelength of 810 nm and a pump wavelength of 750 nm. The energy fluence of the pump pulse at the center of the pump spot is 8~$\mu$J/cm$^2$. (b) Differential reflection single near zero probe delays with pump fluences of (from bottom to top) 10, 20, 40, and 55~$\mu$J/cm$^2$, respectively. (c) Peak differential reflection signal as a function of the injected exciton density. The solid line is a fit. (d) Peak differential reflection signal as a function of the probe wavelength (squares, left axis). The solid line is a photoluminescence spectrum of sample. }
\label{fig:DRoR}
\end{figure}

Figure \ref{fig:DRoR}(a) shows a differential reflection signal of the 810-nm probe pulse as a function of the probe delay (defined as the time delay of the probe pulse with respect to the pump). The differential reflection is defined as the relative change of the reflection of the probe, $\Delta R / R_0 = ( R - R_0) /R_0$, where $R$ and $R_0$ are reflection of the probe with and without the presence of the pump pulse, respectively. We find that the differential reflection signal decays quickly in the first 50 ps and then slowly over several hundred ps. Since the differential reflection is related to the exciton density, its decay reflects the excitonic dynamics. 

In order to establish a precise relation between the differential reflection and the exciton density, we repeat the measurement with different pump fluences. Figure \ref{fig:DRoR}(b) shows the measured differential reflection near zero probe delay with four different pump fluences. In each case, the rising time of the signal is limited by the instrument response. Hence, the excitons injected by the pump pulse instantaneously change the probe reflection. From the pump fluence, we can estimate the injected exciton density by using an absorption coefficient of $2 \times 10^{5}$/cm\cite{jpc12881} and assuming that every pump photon absorbed excites one exciton. Since the exciton lifetime, indicated by the decay of the signal, is much longer than the rising time, we can ignore the decay of the exciton density during the pump pulse and assume that the exciton density at the peak time equals to the injected density. This procedure allows us to related the differential reflection signal to the exciton density, as we plot in Fig. \ref{fig:DRoR}(c). We find that the relation can be accurately described by a saturable absorption model,\cite{bookBoyd}
\begin{equation}
\frac{\Delta R}{R_0} = A \frac{N}{N+N_s},
\label{eq:sa}
\end{equation}
where $A$, $N$, and $N_s$ are a dimensionless constant, the exciton density, and the saturation density, respectively. The solid line in Fig. \ref{fig:DRoR}(c) indicates a fit to the data, with $N_s = (5.8 \pm 0.5) \times 10^{12}$/cm$^2$. Such a saturation density corresponds to an average exciton distance of about 4 nm.  

To further study the mechanism of the saturable absorption, we repeat the measurement at different probe wavelengths, with a pump wavelength of 750 nm and a pump fluence of 4~$\mu$J/cm$^2$. The peak $\Delta R/R_0$ is plotted as the squares in Fig. \ref{fig:DRoR}(d) (left axis). Clearly, the spectrum of $\Delta R/R_0$ coincides with the photoluminescence (PL) spectrum [solid line in Fig. \ref{fig:DRoR}(d)], which is measured with a 633-nm continuous-wave laser excitation. This indicates that the pump-injected excitons reduce the exciton transition strength, but do not broaden nor shift the transition. Such a spectral feature indicates that the dominate mechanism of the absorption saturation is the phase-state filling effect, which often dominates the excitonic nonlinearities in semiconductors.\cite{b326601} However, this is quite different from monolayers of MoS$_2$, in which previous studies have shown rather complicated spectra of transient absorption.\cite{acsnano71072,b88075434} We note that the probe wavelength range in this measurement is limited by the instruments.

The observed excitonic absorption saturation and the unusually large exciton binding energy indicate potential applications of MoSe$_2$ monolayers as saturable absorbers for various nonlinear photonic devices.\cite{afm193077} Here, however, our purpose is to use the absorption saturation to study excitonic dynamics. In principle, the solid line in Fig. \ref{fig:DRoR}(c) allows us to precisely convert the measured $\Delta R/R_0$ to $N$. In this study, however, most measurements are performed with $N \ll N_s$, so that the $\Delta R/R_0$ is approximately proportional to $N$.

\begin{figure}
  \centering
  \includegraphics[width=8.5cm]{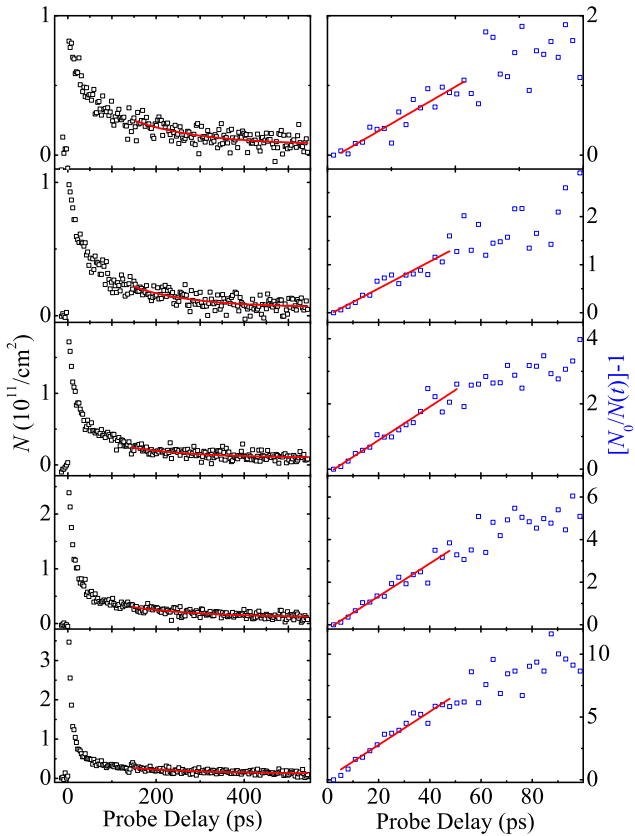}
  \caption{Left column: Exciton density, deduced from the measured differential reflection signal, as a function of the probe delay with different injected densities. The red lines are single exponential fits to the data after 150 ps. Right column: the quantity $N_0/N(t)-1$ calculated from data in left column as a function of the probe delay. The red lines are linear fits.}
  \label{fig:XX}
\end{figure}

We study exciton dynamics at different injection levels. The left column of Fig. \ref{fig:XX} shows that the decay of the exciton density depends strongly on the initially injected density. 
When increasing the injected density, a fast decay component develops. Such a density-dependent decay is not expected from a non-interacting exciton system, thus indicating strong exciton-exciton interactions.

It is well known that in strongly confined systems, such as organic crystals, excitation of nearby molecules can result in annihilation of excitons due to their strong interactions.\cite{am121655,am14701,b11716} The exciton-exciton annihilation has also been observed in one-dimensional structures, such as semiconducting carbon nanotubes.\cite{np554,l94157402} However, its observation in quasi 2D systems, such as semiconductor quantum wells, is rare. Since MX$_2$ monolayers are atomically thin, strong exciton-exciton coupling can be expected. Including exciton-exciton annihilation, the rate equation of the exciton density can be written as
\begin{equation}
\frac{dN}{dt}=-\frac{1}{\tau}N-\frac{1}{2} \gamma N^2,
\label{eq:rate}
\end{equation}
where $\tau$ and $\gamma$ are the exciton lifetime and exciton-exciton annihilation rate, respectively.\cite{l94157402} One could attempt to compare the solution of this equation with the data. However, for pedagogical considerations, here we discuss separately contributions of the two mechanisms. This is possible because the exciton-exciton annihilation is only significant in early times when the densities are high, while the single-particle process dominates decays on longer time scales with lower densities. In fact, we find that the data after 150 ps can be satisfactorily fit by a single exponential function, as indicated by the red lines in Fig. \ref{fig:XX} (left column).

\begin{figure}
  \centering
  \includegraphics[width=8.5cm]{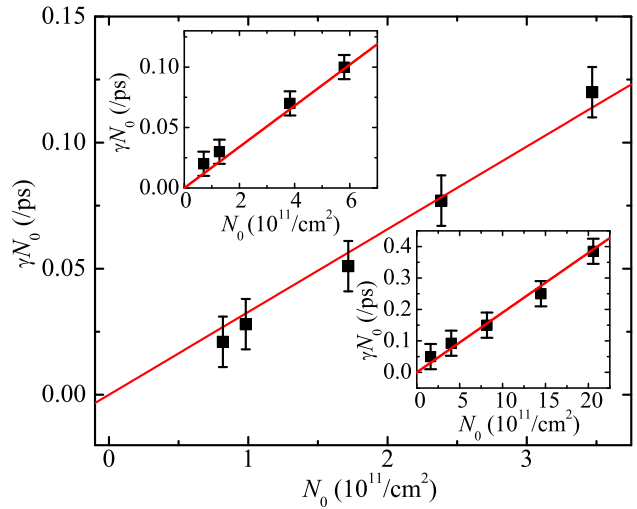}
  \caption{Rate of increase of the quantity $N_0/N(t)-1$, deduced from linear fits shown in the right column of Fig. \ref{fig:XX}, as a function of injected exciton density. The solid line indicate a linear fit. The two insets show similar results from two other samples.}
    \label{fig:slope}
\end{figure}

Without the first term on the right hand side, the solution to Eq. \ref{eq:rate} is simply
\begin{equation}
\frac{N_0}{N(t)} - 1 =  \gamma N_0 t,
\label{eq:XX}
\end{equation}
where $N_0$ is the initially injected exciton density at $t=0$. In viewing of this, we calculate $N_0/N(t)-1$ from the data shown in each panel in the left column of Fig. \ref{fig:XX}, and plot it as a function of $t$ in the right column. In the first 50 ps, the data are consistent with Eq. \ref{eq:XX}, as indicated by the solid lines. We attribute the deviation from linear after 50 ps to the contribution of the first term in Eq. \ref{eq:rate}. From linear fits, shown as the solid lines in Fig. \ref{fig:XX} (right colume), we deduce the slopes ($\gamma N_0$). We find that the slope is indeed proportional to $N_0$, as shown in Fig. \ref{fig:slope}. From a linear fit, shown as the solid line in Fig. \ref{fig:slope}, we obtain an exciton-exciton annihilation rate of $\gamma = 0.33 \pm 0.06$ cm$^2$/s. We repeat the measurement on two other monolayer samples, and obtained similar results, as summarized in the two insets of Fig. \ref{fig:slope}. The observation of the exciton-exciton annihilation illustrates the strong interaction between excitons in monolayers of MoSe$_2$. In a recent study, strong inter-exciton coupling in monolayers of MoS$_2$ was revealed, although exciton-exciton annihilation was not observed.\cite{b88075434}

\begin{figure}
  \centering
  \includegraphics[width=8.5cm]{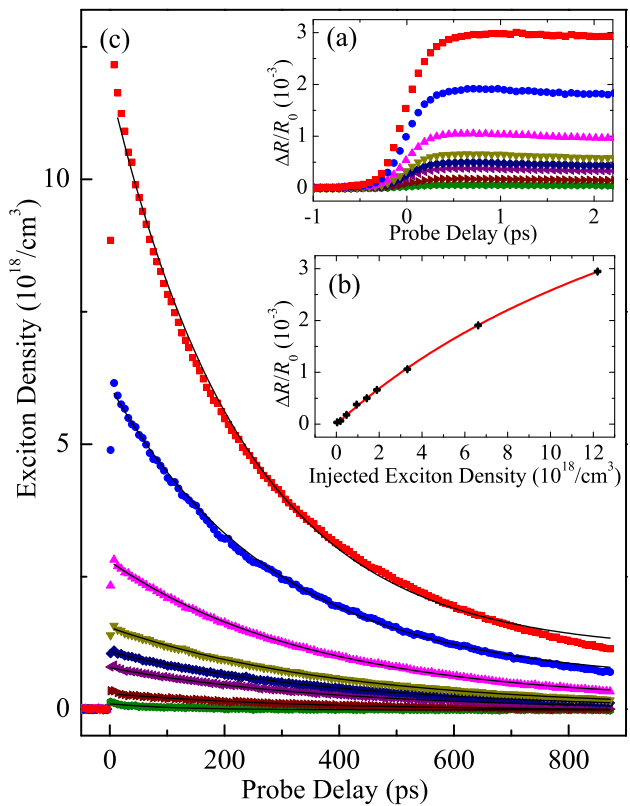}
  \caption{Exciton dynamics in bulk MoSe$_2$. (a) Differential reflection signal with the probe delays in a short time range near zero delay measured from a bulk MoSe$_2$ sample. The pump fluences are (from top to bottom) 21.8, 11.8, 5.9, 3.3, 2.5, 1.7, 0.8, and 0.3 $\mu$J/cm$^2$, respectively. (b) Peak differential reflection signal as a function of the injected peak exciton density. The red line is a fit to the data using Eq.~\ref{eq:sa}. (c) Decay of exciton density at different initial injection levels. The solid lines are exponential fits to the data. }
    \label{fig:bulk}
\end{figure}

For comparison, we also study the exciton dynamics in a bulk MoSe$_2$ sample, which is fabricated from the same crystal used to make the monolayers. The measurement is performed with the same setup. Excitons are injected by a 750-nm pump pulse, and probed with an 810-nm pulse.  Figure \ref{fig:bulk}(a) shows the differential reflection signal measured in a short time range of a few ps, with different pump fluences. As we observed in monolayers, the differential reflection signal rise to a peak quickly, limited by the time resolution. This indicates the instantaneous phase-state filling effect of the resonantly injected excitons. From these data, we obtain the peak differential reflection signal as a function of the injected exciton density, as shown in Fig. \ref{fig:bulk}(b). Here, the bulk exciton density represents its peak value at the center of the pump spot and at the same surface, and is deduced from the pump fluence. Similar to the monolayers, the differential reflection signal can be well described by the saturable absorption model (Eq. \ref{eq:sa}), as indicated as the solid line. We obtain a saturation density of $(2.2 \pm 0.3) \times 10^{19}$/cm$^3$.

Next, we measure the differential reflection signal over a longer time range of about 1 ns with various pump fluence, and deduce the exciton density by using the solid line shown in Fig. \ref{fig:bulk}(b). The results are plotted in Fig. \ref{fig:bulk}(c). No signature of exciton-exciton annihilation is observed, and all the data can be satisfactorily fit by single exponential functions with time constants in the range of 300 - 400 ps, as indicated as the solid lines in Fig. \ref{fig:bulk}(c). We note that the highest density of $1.2 \times 10^{19}$ /cm$^3$ used in this measurement corresponds to an areal density of $8.4 \times 10^{11}$ /cm$^2$ in the first atomic layer (0.7-nm thick), which is larger than areal densities used in the monolayer measurement. Finally, we attribute the longer decay time of the exciton density in bulk sample to longer exciton lifetimes in bulk, probably due to the indirect bandgap and less surface contributions to the exciton recombination.

In summary, we studied excitonic dynamics in MoSe$_2$ by femtosecond transient absorption microscopy and observed exciton-exciton annihilation in monolayers, which is absent in bulk under similar conditions. This process reveals strong coupling between excitons in this strongly confined two-dimensional system. We also found that the exciton density decay time is about twice longer in bulk than monolayers. Furthermore, we observed saturation absorption in both monolayer and bulk, and deduced saturation densities. This observation, combined with the unusually large exciton binding energies, suggest potential uses of MoSe$_2$ monolayers and bulk as saturable absorbers. 

We acknowledges support from the US National Science Foundation under Awards No. DMR-0954486, the National Basic Research Program 973 of China (2011CB932700, 2011CB932703), Chinese Natural Science Fund Project (61077044), and Beijing Natural Science Fund Project (4132031). 
 
%\bibliography{/users/huizhao/Documents/Bibfile/literature.bib}

\end{document}